\documentclass{PoS}

\title{Exclusive scalar \mbox{\boldmath $f_0(1500)$} meson production}

\ShortTitle{Exclusive scalar \mbox{\boldmath $f_0(1500)$} meson production}

\author{\speaker{Piotr ~Lebiedowicz} $^a$ and Antoni Szczurek $^{ab}$\\
       \llap{$^a$} Institute of Nuclear Physics PAS, PL-31-342 Cracow, Poland\\
       \llap{$^b$} University of Rzesz\'ow, PL-35-959 Rzesz\'ow, Poland\\    
       E-mail: \email{piotr.lebiedowicz@ifj.edu.pl},
               \email{antoni.szczurek@ifj.edu.pl}}

\abstract{We evaluate differential distributions for exclusive scalar
$f_0(1500)$ meson (glueball candidate) production.
Both QCD diffractive, 
pion-pion meson exchange current (MEC) components as well as 
double-diffractive mechanism with intermediate pionic loop 
are calculated. 
The pion-pion component, which can be reliably calculated, 
dominates close to the threshold 
while the diffractive component may take over only for larger energies.
At the moment only upper limit for the QCD-diffractive component
can be obtained. 
The diffractive component is calculated based on 
the Khoze-Martin-Ryskin approach proposed for diffractive 
Higgs boson production.
Different unintegrated gluon distribution functions (UGDFs) from 
the literature are used.  
Rather large cross sections (due to pion-pion fusion) are predicted 
for PANDA energies, where the gluonic mechanism is shown 
to be negligible.}

\FullConference{European Physical Society Europhysics Conference on High Energy Physics\\
          July 16-22, 2009\\
          Krakow, Poland}

\begin{document}

\section{Introduction}

Many theoretical calculations, including lattice QCD,
predicted existence of glueballs (particles dominantly made of 
gluons) with masses $M >$ 1.5 GeV. 
No one of them was up to now unambiguously identified.
The nature of scalar mesons below 2 GeV is also not well understood. 
The lowest mass meson considered as a glueball candidate is 
a scalar $f_0(1500)$ \cite{AC96} observed by the Crystall Barrel
Collaboration in $p\bar p$ annihilation 
\cite{Crystall_Barrel_f0_1500}.
It was next confirmed by the WA102 Collaboration in central
production in $pp$ collisions
in two-pion \cite{WA102_f0_1500_2pi} and four-pion \cite{WA102_f0_1500_4pi} decay channels.

\section{Mechanisms of exclusive scalar $f_0(1500)$ meson production}
\begin{figure}[!htb]    
\begin{center}
(a)\includegraphics[width=0.25\textwidth]{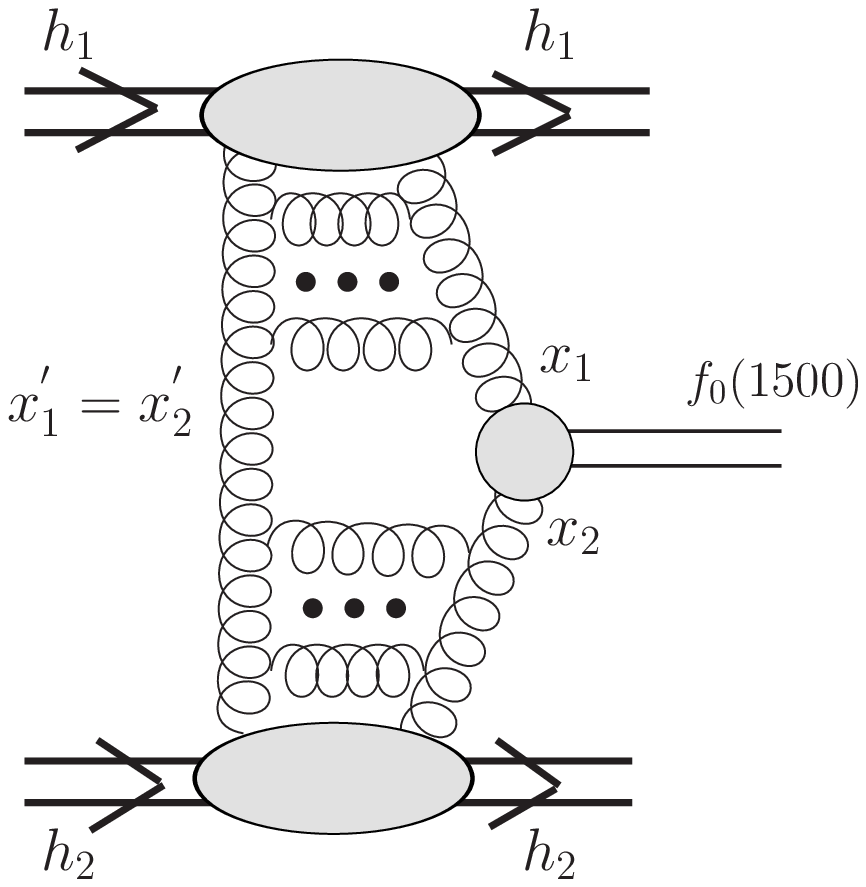}
(b)\includegraphics[width=0.25\textwidth]{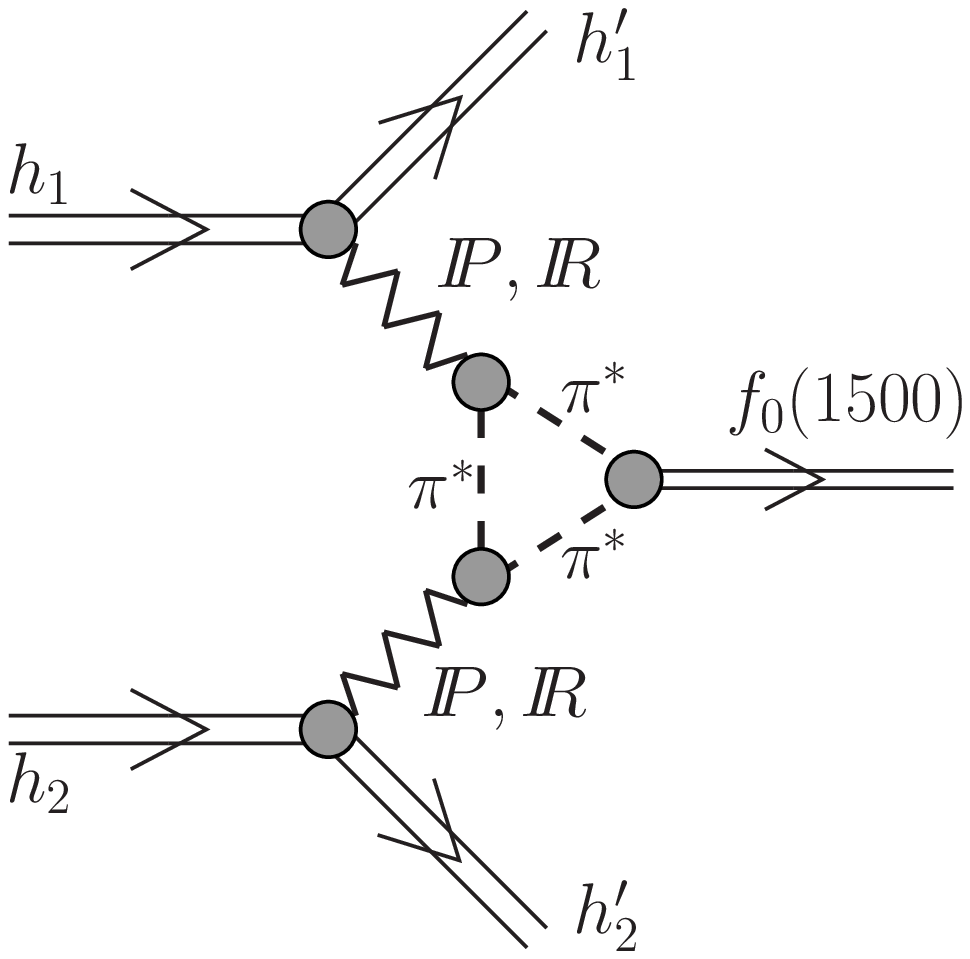}
(c)\includegraphics[width=0.21\textwidth]{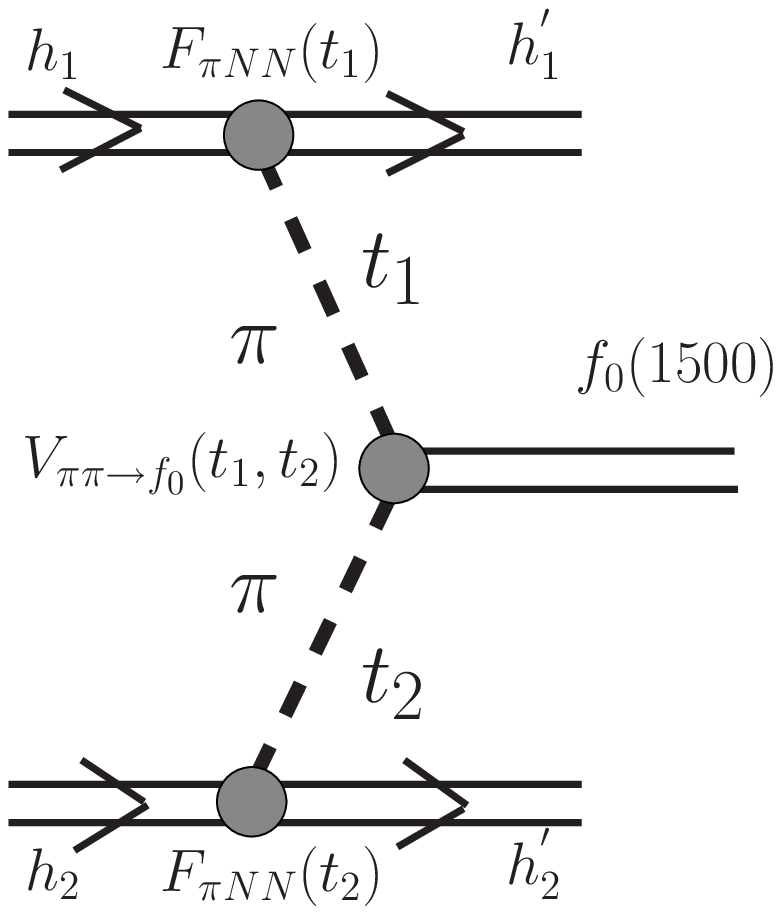}
\end{center}
   \caption{\label{fig:mechanisms}
   \small  The sketch of the bare mechanisms of exclusive scalar
$f_0(1500)$ meson production:
(a) the QCD mechanism, 
(b) double-diffractive mechanism with intermediate pionic triangle 
and (c) pion-pion fusion. 
}
\end{figure}

We concentrate on exclusive production of scalar $f_0(1500)$ 
in the following reactions:
$p p \to p p f_0(1500)$,
$p \bar p \to p \bar p f_0(1500)$,
$p \bar p \to n \bar n f_0(1500)$.
While the first process can be measured at the J-PARC complex,
the latter two reactions could be measured by the PANDA Collaboration.
We have proposed a different mechanisms (shown in Fig.\ref{fig:mechanisms}) 
for exclusive scalar $f_0(1500)$ meson production (more details can be found in Ref. \cite{SL09}).

If $f_0(1500)$ is a glueball (or has a strong glueball component \cite{CZ05})
then the diffractive mechanism (see Fig.\ref{fig:mechanisms}a)
may be important.
This mechanism is often considered as the dominant mechanism of 
exclusive Higgs boson \cite{KMR}
and $\chi_c(0^+)$ meson \cite{PST07} production at high energies.
At lower energies ($\sqrt{s} <$ 20 GeV) other processes may 
become important as well.
Since the two-pion channel is one of the dominant decay channels
of $f_0(1500)$ (34.9 $\pm$ 2.3 \%) \cite{PDG}
one may expect the two-pion fusion (see Fig.\ref{fig:mechanisms}c)
to be one of the dominant mechanisms at the FAIR energies. The two-pion fusion can be
relative reliably calculated in the framework of 
meson-exchange theory. The pion coupling to the nucleon is well
known \cite{Ericson-Weise}.

\section{Results and Conclusions}

\begin{figure}[!htb]    
\begin{center}
	\includegraphics[width=0.29\textwidth]{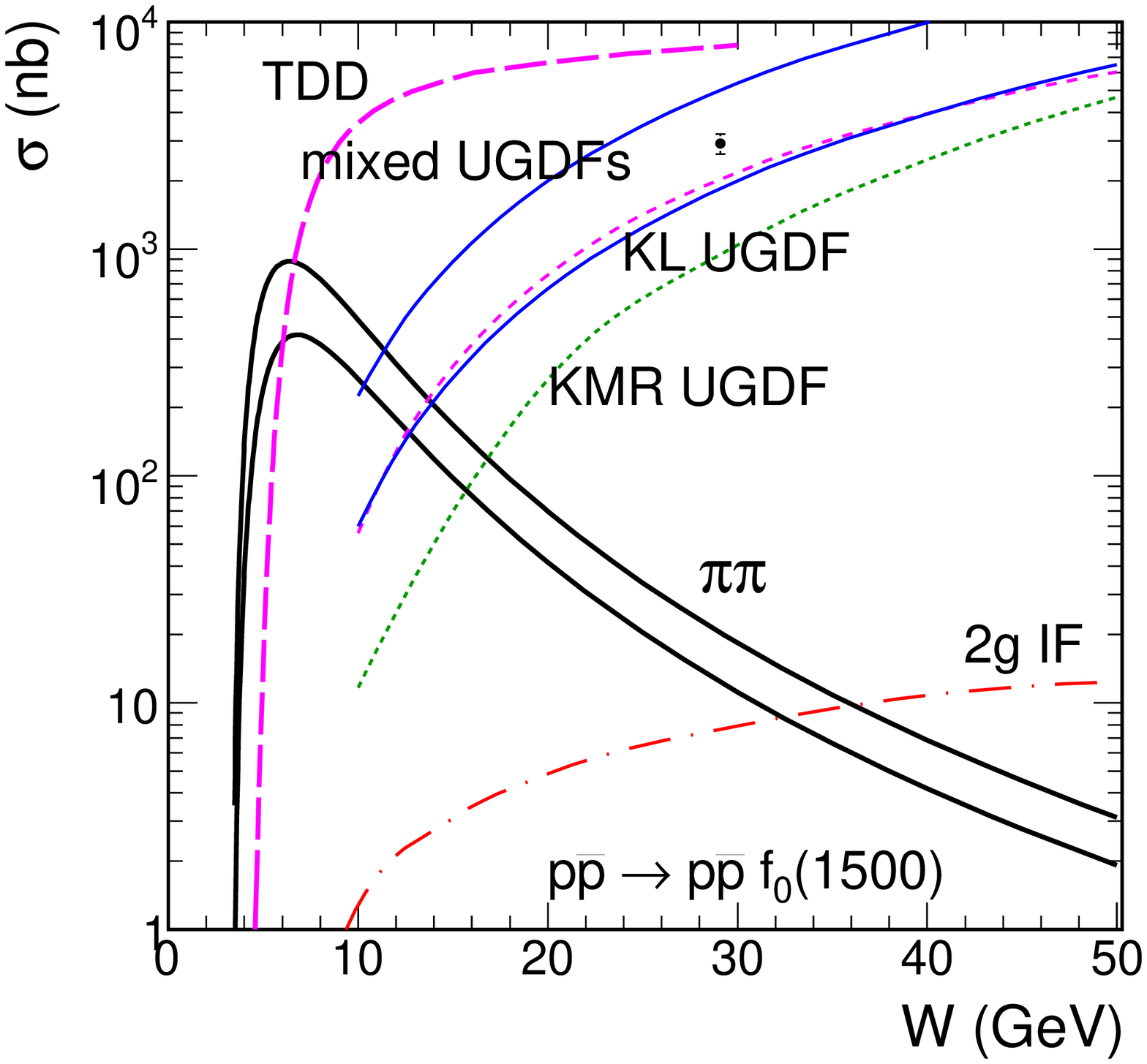}
\space \space \space
	\includegraphics[width=0.29\textwidth]{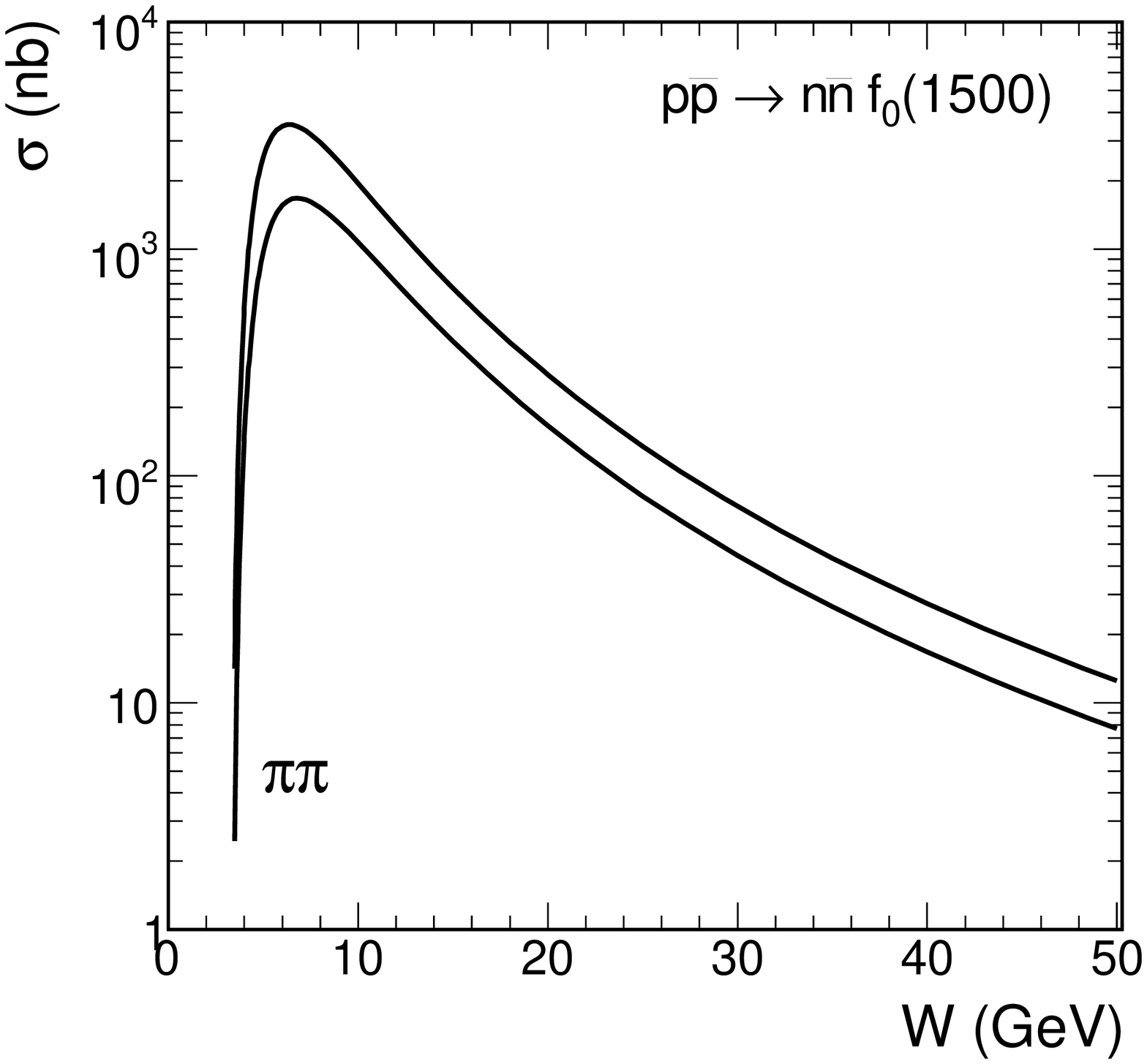}
   \caption{\label{fig:sigma_W}
   \small The integrated cross section as a function
of the center-of-mass
energy for $p \bar p \to p \bar p f_0(1500)$ (left panel)
and $p \bar p \to n \bar n f_0(1500)$ (right panel) reactions.
The thick solid lines: $\pi\pi$ contribution,
the dashed line: QCD diffractive contribution (KL UGDF), 
the dotted line: the KMR approach,
the thin solid lines: "mixed" UGDF (KL $\otimes$ Gauss) and
the long-dashed line: the mechanism with intermediate pionic triangle.
The WA102 experimental data point at $W = 29.1$ GeV is from \cite{kirk}.
}
\end{center}
\end{figure}

\begin{figure}[!htb]    
(a)\includegraphics[width=0.29\textwidth]{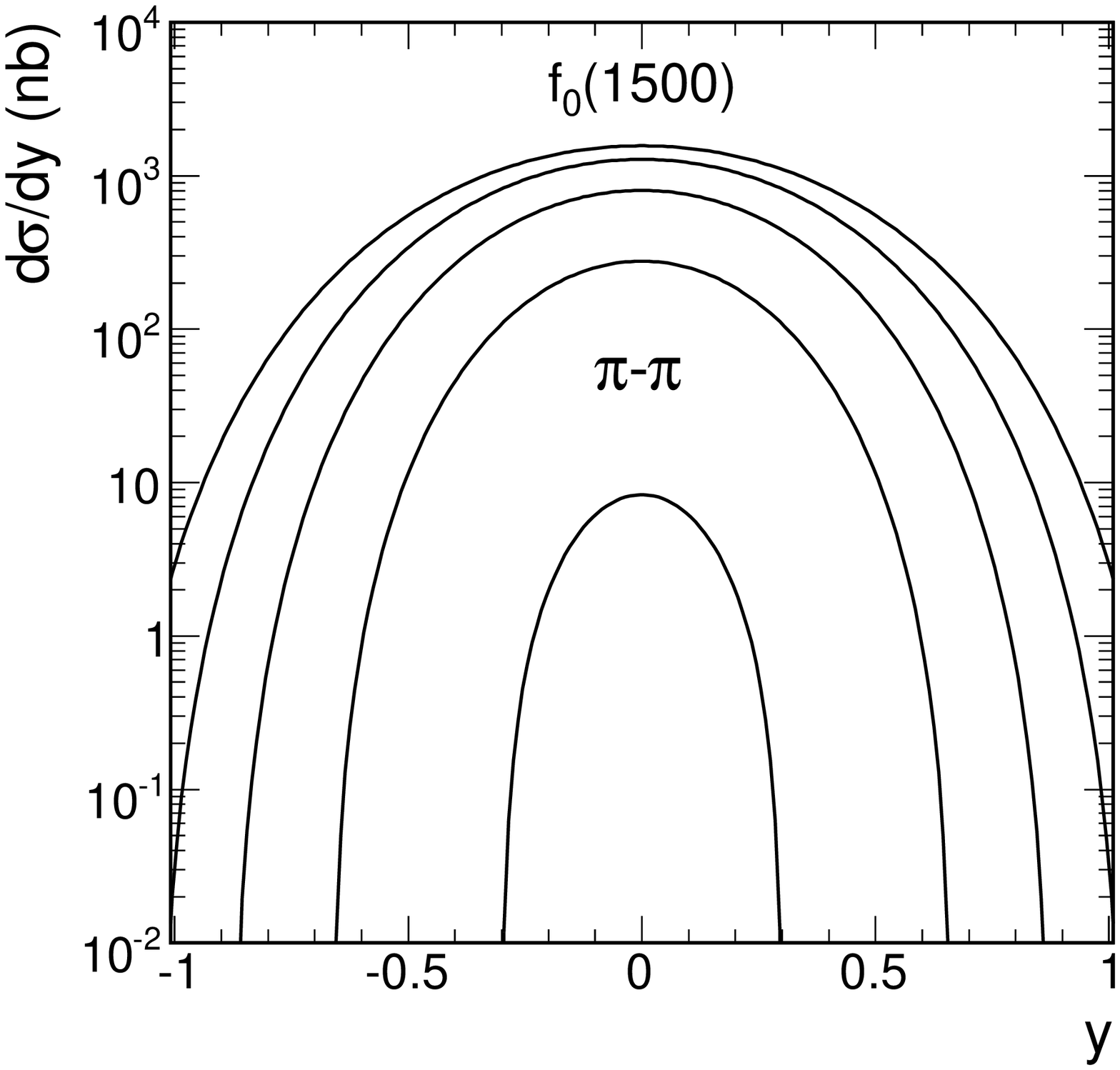}
(b)\includegraphics[width=0.29\textwidth]{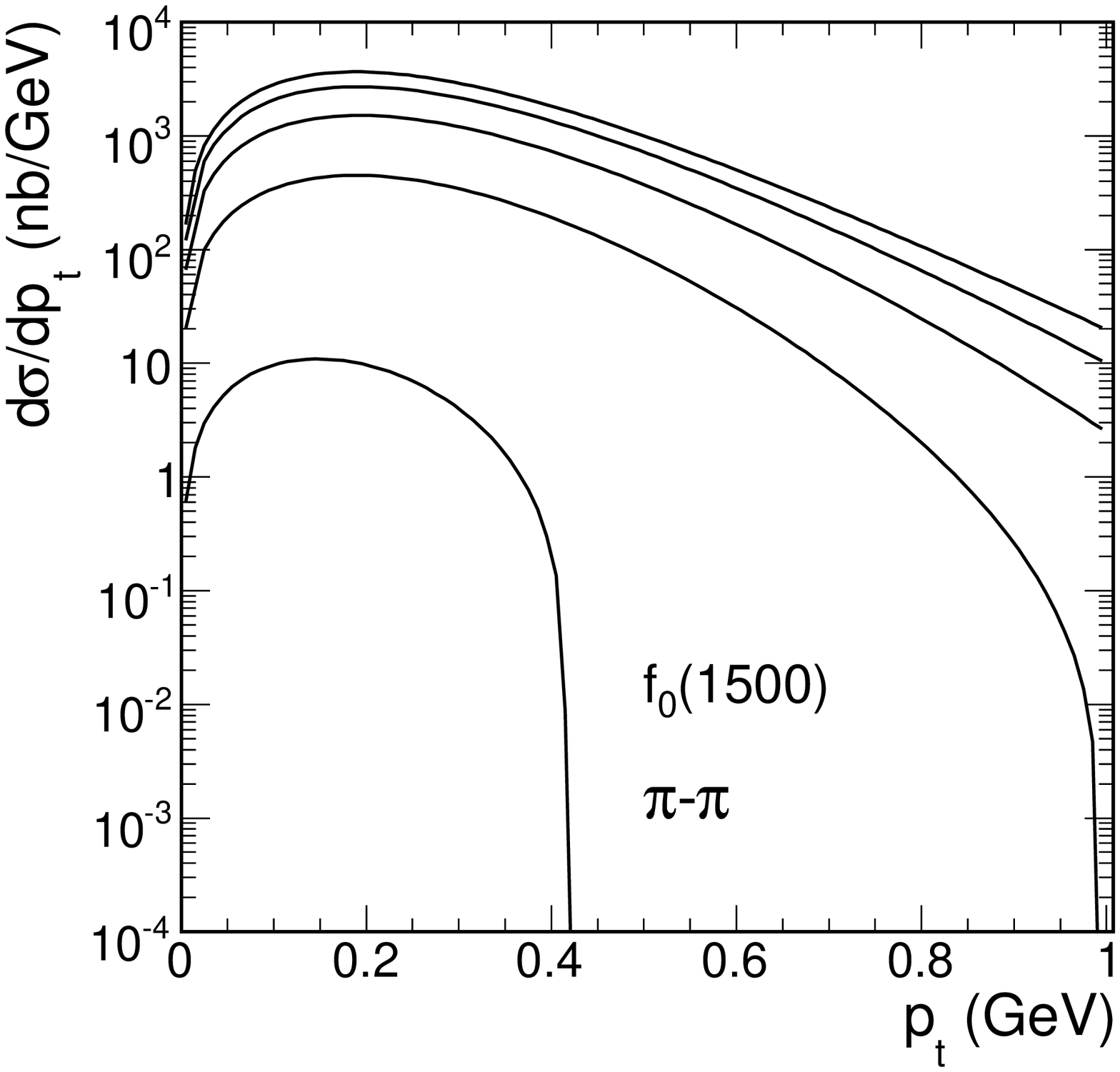}
(c)\includegraphics[width=0.29\textwidth]{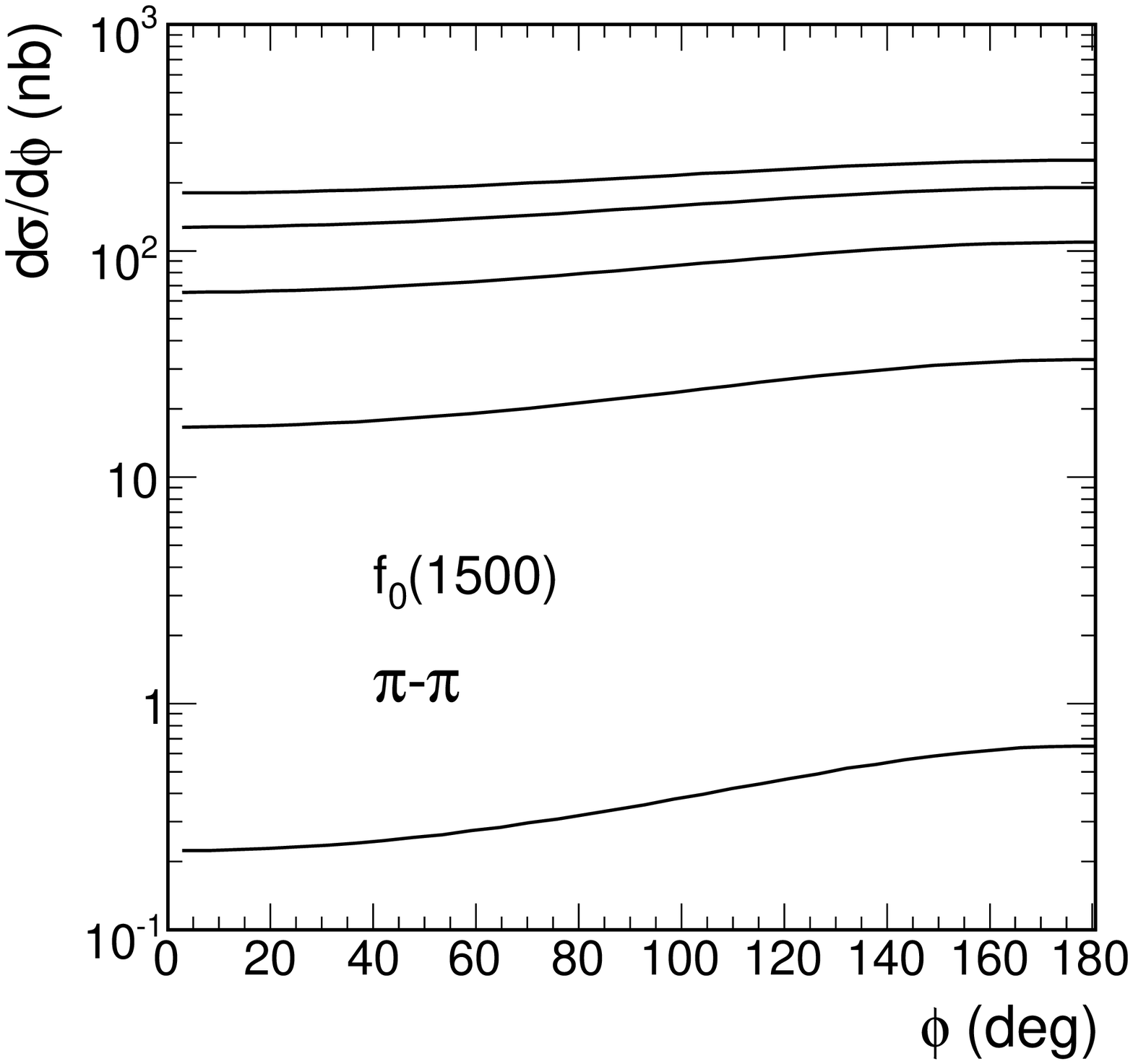}
   \caption{\label{fig:dsig_pipi}
   \small
Differential cross sections
$\frac{d\sigma}{dy}$ (a), $\frac{d\sigma}{dp_{t}}$ (b) and $\frac{d\sigma}{d\phi}$ (c)
for the reaction $p \bar p \to n \bar n f_0(1500)$ ($\pi^+ \pi^-$ fusion only)
at $W$ = 3.5, 4.0, 4.5, 5.0, 5.5 GeV (from bottom to top).
}
\end{figure}

We have estimated the integrated cross 
section for the exclusive $f_0(1500)$ meson production (see Fig.\ref{fig:sigma_W}).
We have included both gluon induced diffractive and triangle-double-diffractive
mechanisms as well as the $\pi\pi$ exchange contributions. 
We predict the dominance of the $\pi\pi$ contribution
close to the threshold and the dominance
the diffractive components at higher energies.
In Fig.\ref{fig:dsig_pipi} we present differential cross sections for the $\pi\pi$ exchange mechanism at energies of future experiments at HESR at 
the FAIR facility in GSI.

The experimental studies of exclusive production of $f_0(1500)$ are 
not easy at all as in the $\pi \pi$ decay channel one
expects a large continuum. We have performed an involved calculation
of the four-body $p \bar p \pi^+ \pi^-$ background.
Our calculation \cite{SL09} shows that imposing extra cuts should allow to extract
the signal of the glueball $f_0(1500)$ candidate at the highest PANDA 
energy.


\begin{thebibliography}{99}
\bibitem{AC96}
C. Amsler and F.E. Close, Phys. Rev. {\bf D53} (1996) 295;
F.E. Close, Acta Phys.Polon. {\bf B31} (2000) 2557.

\bibitem{Crystall_Barrel_f0_1500}
C. Amsler et al., Phys. Lett. {\bf B342} (1995) 433;
C. Amsler et al., Phys. Lett. {\bf B353} (1995) 571;
C. Amsler et al., Phys. Lett. {\bf B380} (1996) 453;
A. Abele et al., Phys. Lett. {\bf B385} (1996) 425;


\bibitem{WA102_f0_1500_2pi}
D. Barberis et al. (WA102 Collaboration), Phys. Lett. {\bf B462} (1999) 279.

\bibitem{WA102_f0_1500_4pi}
D. Barberis et al. (WA102 Collaboration), [{\tt hep-ex/0001017}].

\bibitem{SL09}
A. Szczurek and P. Lebiedowicz, Nucl. Phys. {\bf A826} (2009) 101, [{\tt nucl-th/0906.0286}].



\bibitem{CZ05}
F.E. Close and Q. Zhao, Phys. Rev. {\bf D71} (2005) 094022.

\bibitem{KMR}
V.A. Khoze, A.D. Martin and M.G. Ryskin, Phys. Lett. {\bf B401} (1997) 330;
V.A. Khoze, A.D. Martin and M.G. Ryskin, Eur. Phys. J. {\bf C23} (2002) 311;


\bibitem{PST07}
R.~S.~Pasechnik, A.~Szczurek and O.~V.~Teryaev, Phys. Rev. {\bf D78} (2008) 014007.


\bibitem{PDG}
C. Amsler \textit{et al}. (Particle Data Group), Phys. Lett. {\bf B 667} (2008) 1.

\bibitem{Ericson-Weise}
T. Ericson and A. Thomas, \emph{Pions and Nuclei}, Oxford University Press, 1988.





\bibitem{kirk}
A. Kirk, Phys. Lett. {\bf B489} (2000) 29.


\end{thebibliography}
\end{document}